\documentclass[12pt]{article}
\usepackage{amsmath,amssymb,amsfonts,amsbsy}

\usepackage{cite}

\usepackage[pdftex, 
colorlinks,
pdfauthor=Sergio,
]{hyperref}

\usepackage{geometry}                % See geometry.pdf to learn the layout options. There are lots.
\usepackage{graphicx}
\usepackage{epstopdf}
\usepackage{slashed}

\newcommand{\be}{\begin{equation}}
\newcommand{\ee}{\end{equation}}

\newcommand{\reff}[1]{(\ref{#1})}
\newcommand{\tr}{\mathop{\rm tr}\nolimits}
\newcommand{\bone}{1\!\!1}

\title{Transfer matrix for Kogut-Susskind fermions\\ in the spin basis}
\author{
  {\small Sergio Caracciolo}                       \\[-1.7mm]
  {\small\it Dipartimento di Fisica and INFN}      \\[-1.7mm]
  {\small\it Universit\`a degli Studi di Milano}   \\[-1.7mm]
  {\small\it via Celoria 16, I-20133 Milano, ITALY}                \\[-1.7mm]
  {\small\tt Sergio.Caracciolo@mi.infn.it},        \\[-1.7mm]
  {\protect\makebox[5in]{\quad}}
   \\
  {\small Fabrizio Palumbo}  \\[-1.7mm]
  {\small\it INFN -- Laboratori Nazionali di Frascati}      \\[-1.7mm]
  {\small\it P.~O.~Box 13, I-00044 Frascati, ITALY}       \\[-1.7mm]
   {\small\tt fabrizio.palumbo@lnf.infn.it}       \\[-1.7mm]
  \\
 {\protect\makebox[5in]{\quad}} }

\begin{document}
\maketitle

\thispagestyle{empty}

\begin{abstract}

In the absence of interaction it is well known that the  Kogut-Susskind regularizations of  fermions in the spin and flavor basis are equivalent to each other.
In this paper we clarify the difference between  the two formulations in the presence of interaction with  gauge fields.
We then derive an explicit expression of the transfer matrix in the spin basis by a unitary transformation on that one in the flavor basis which is known.
The essential key ingredient is the explicit construction of the fermion Fock space for variables which live on blocks. 
Therefore the transfer matrix generates  time translations of two lattice units.
\end{abstract}

\clearpage

\vfill\eject

\section{Introduction}

The naive discretization of the Dirac equation on the lattice~\cite{Wilson74} leads to a replication of the fermionic states, known as lattice fermion {\em doubling}~\cite{Wilson75}.
The doublers appear as spurious poles in the fermion propagator at the nonzero corners of the Brillouin zone. 
The Wilson way of  removing the doublers is to give them a mass which becomes infinite in the continuum limit, at the cost of an explicit breaking of chiral invariance on the lattice~\cite{Wilson75}.

In the Kogut-Susskind~\cite{KS, BKS, SS76, Susskind} lattice formulation for relativistic fermions~\cite[Chap.~4]{MM} the doublers are instead interpreted as physical fields by the introduction of additional quantum numbers. This has been done in two ways.
In the former approach, first the fermion field is  reduced to a single component per site by a procedure called {\em spin diagonalization}, and, for this reason, this method is referred to as the one in the {\it spin basis}.  Afterwards  
spin and flavour degrees of freedom are associated to different corners of an elementary hypercube on the lattice~\cite{Kluber81, Kluber83, Gliozzi}, and therefore sometimes  fermions in this formulation are said to be {\it staggered}.
In the latter approach~\cite{STW, Vandendoel, Golterman}, said in the {\em flavour basis}, the additional quantum numbers, called {\em taste}, are associated, together with the  spin, with blocks corresponding to the hypercubes of the spin basis of size twice the lattice spacing.

%The same continuum Lagrangian takes different forms when regulated in the two bases.
In the absence of coupling with gauge fields  these forms  are changed into one another by a linear transformation on the fermion fields, but in the presence of gauge fields they are not equivalent, as we shall make clear in the following.  Their difference is of consequence in the construction of the corresponding transfer matrices.  

For Kogut-Susskind fermions in the flavour basis a simple operator realization of the transfer matrix is known~\cite{Fab02}. It has been built in close analogy with the case of Wilson fermions~\cite{LuscherTM, Creutz77, Creutz87, Smit, Creutz99} (see also~\cite{MP}), the only difference being that it performs time translations by one block  instead of one lattice spacing. 

The situation is more complex for Kogut-Susskind fermions in the spin basis~\cite{STW, Vandendoel, Banks}, because all attempts at constructing a positive definite transfer matrix that performs time translations by a single lattice spacing failed.  The difficulty was circumvented by looking at time translations by two lattice spacings. 
Here we meet with a subtlety. We must distinguish whether the Fock space is built on one or two  time slices. 
In the first case, the square root of the matrix which translates by two lattice spacings is the one that translates by one lattice spacing.  In the second case, instead, translations by one lattice spacing are not defined at all.
This seems to be the case with Kogut-Susskind fermions, but the necessary construction of the Fock space on blocks, in the spin basis, has not been made explicit. 

%This seems to imply a Fock space built  on blocks, but such a construction has not been explicitly shown.  A minor difficulty might  originate from the fact that in the reduced Lagrangian does not exist a mass term defined on a single site.
 
We became interested  in a formulation of the transfer matrix in the spin basis in the  framework of relativistic field theories of
fermions whose partition function is dominated by bosonic composites~\cite{CLP}. This subject  became  for us more relevant in the development of  an approach to  QCD hadronization (meant as the replacement of the QCD degrees of freedom by hadronic ones) that makes use of the operator form of the transfer matrix~\cite{ Palu, CPV, noi-prd, diquarks}. Using Kogut-Susskind fermions, because of the lack of a convenient formulation of the transfer matrix, we were able to express our results only in the flavour basis. Numerical simulations are, instead, usually performed in the spin basis, because they are much faster.  We were thus motivated to find an operator form of the transfer matrix in this latter basis as well. % , such as to make comparisons with standard approaches using the functional form easy.   
Since apparently in any case we should resign to time translations by one block, we decided to get an expression of the transfer matrix in the spin basis by a linear transformation from the flavour basis.

We deem that the question might be of more general interest, and therefore we report our results in the present paper.  
In Sect.~\ref{ks} we remind for the convenience of the reader and  in order to establish the notation  what is relevant for the following about the Kogut-Susskind regularization.  We adopt the notations of Montvay and M\"{u}nster~\cite{MM} with some minor changes that will be specified. 
In Sect.~\ref{tl} we perform the transformation of the action from the flavour to the spin basis. Most of the results, with some qualification, are well known, but we think this Section is a necessary preparation for Sect.~\ref{ttm}, in which  we perform the transformation of the transfer matrix.

\section{Kogut-Susskind fermions}\label{ks}

%Kogut-Susskind fermion fields  are defined on the sites of  a basic cubic lattice or on hypercubic blocks built on the basic lattice. 
Let  $x_{\mu}$ be the coordinates of hypercubic lattice sites,  $0\leq x_\mu \leq L_\mu-1$ , $0\leq \mu\leq 3$ (Montvay and M\"{u}nster in~\cite{MM} use indices from 1 to 4), and $y_\mu$   the coordinates of  hypercubic blocks. They are related by
\be
x_\mu = 2 y_\mu + \eta_\mu 
\ee
with  $0 \leq y_\mu \leq L'_\mu -1$, $L_\mu = 2 L'_\mu$, and $\eta_\mu = 0,1$  the position vectors within the block.  The sum over lattice points can be split into the sum over the blocks and the sum over the sites within a block, that is
\be
\sum_x = \sum_y  \sum_\eta \,.
\ee
We denote by $\psi_x$ the fermionic fields on the lattice sites,  and by $q_{y}^{\alpha a}$ the fields on the blocks. The latter have Dirac  spinor indices $1\leq \alpha \leq 4$, in greek letters, and taste indices $1\leq a\leq 4$, in latin letters. 

It is important to remark that the gauge transformations in the first case act at the sites of the basic lattice, in the second at the coordinates of the blocks
\be
\psi_x \,  \to  \, g_x \,  \psi_x \, , \qquad \, q_{y}^{\alpha a}\,  \to \, g_y \,q_{y}^{\alpha a}\, .
\ee
While $g_y$ is the same transformation for all $x $  in a given  block with coordinate $y$,  $g_x$ will in general  change also within the same block.

\subsection{The flavour basis}

The gauge link variables on the blocks are denoted by $U_{\mu}(y)$. Under gauge transformations they change according to the rule
\be
U_{\mu}(y) \, \to \, 
 g_y\,  U_{\mu}(y)\,  g_{y+{\hat \mu}}^{\dagger}\,.
\ee
The action of the fermion fields  in the flavour basis can be written as
\be
S(U) \, = \, 2^4 \sum_y {\mathcal L}_q(U)
\ee
where the factor $16$ keeps into account the volume of the elementary cell when using variables defined on the blocks, and the Lagrangian  in the flavour basis is
\begin{multline}
%{\mathcal L}_q(U)  =  {\overline q}(y)  \left \{     m  \,  (\bone \otimes \bone)
%+ \sum_{\mu=0}^3      \left[  (\gamma_{\mu} \otimes \bone)   \, \frac{1}{2} \left( \nabla_{\mu}^{(+)} + \nabla_{\mu}^{(-)}  \right)
%- (\gamma_5 \otimes  t_5 t_{\mu})   \, \Delta_{\mu} \right] q \right\}_y
{\mathcal L}_q(U)  :=  m  \,  {\overline q}_y        (\bone \otimes \bone) q_y  \\
+ \sum_{\mu=0}^3    {\overline q}_y  \left \{   \left[  (\gamma_{\mu} \otimes \bone)   \, \frac{1}{2} \left( \nabla_{\mu}^{(+)} + \nabla_{\mu}^{(-)}  \right)
- (\gamma_5 \otimes  t_5 t_{\mu})   \, \Delta_{\mu} \right] q \right\}_y
\label{Lflavour}
\end{multline}
the flavour matrices $t_\mu$ are defined for $\mu=0,\dots,3$  and $\mu=5$ by
\be
t_\mu = \gamma_\mu^T := t_\mu^\dagger
\ee
and the other operators are defined in terms of translations on the blocks
\be
\left[ T_{\mu}^{(\pm)} f \right]_y := 2^4 \sum_{y'} \frac{1}{2^4} \, \delta_{y', y \pm  \hat{\mu}} \, f(y') = f(y \pm  \hat{\mu} )\, 
\ee
and the identity on the blocks
\be
\left[ \bone f \right]_y := 2^4 \sum_{y'} \frac{1}{2^4} \, \delta_{y', y} \, f(y) = f(y )\, 
\ee
according to
\begin{align}
 \nabla_{\mu}^{(+)} & := \frac{1}{2} \left( U_{\mu} \,T^{(+)}_{\mu}  - \bone \right) \, , \qquad \, 
 \nabla_{\mu}^{(-)} := \frac{1}{2} \left( \bone - T^{(-)}_{\mu}  U_{\mu}^{\dagger}\right) \label{covder} \\
\Delta_{\mu}  & := \frac{1}{2} \left(  \nabla_{\mu}^{(+)} -  \nabla_{\mu}^{(-)} \right) = \frac{1}{4}   \left( U_{\mu} \,T^{(+)}_{\mu}+T^{(-)}_{\mu}  U_{\mu}^{\dagger} - 2 \, \bone \right)  \, .
\end{align}
We can recognize that the projections of the fermionic field
\be
q_{+}= P_{+}\, q\,, \qquad \, q_{-}^{\dagger} = P_{-}\, q
\ee
where
\be
P_{\pm} = { \frac{1}{2} } ( \bone \otimes \bone \mp \gamma_0 \gamma_5 \otimes t_5 t_0 ) 
\ee
propagate forward/backward in time, and therefore describe particles/antiparticles respectively. Accordingly we introduce  creation and annihilation operators ${\hat q_{\pm}}^{\dagger}, {\hat q_{\pm}}$. They are defined at one and the same time, so that  in addition to spin and flavour, they depend on the spatial position only, denoted by boldface letters. They satisfy canonical anticommutation relations
\be
\{   (   {\hat q_{\pm}}^{\dagger} )_ {{\bf y}_1}^{a \alpha }, ({\hat q_{\pm}})_{{\bf y}_2}^{\beta b }\} = \frac{1}{8} \, \delta_{{\bf y}_1 {\bf y}_2} 
P_\pm^{\beta b, \alpha a } \,,
\qquad \,   \{(   {\hat q_{\pm}}^{\dagger} )_ {{\bf y}_1}^{a \alpha }, ({\hat q_{\mp}})_{{\bf y}_2}^{\beta b}\}=0 \,.
\label{comm}
\ee
As the factor $\frac{1}{8}$  accounts for the spatial volume of the blocks,  the above anticommutation relations become canonical in the basis in which $P_{\pm}$ are diagonal.

The transfer matrix corresponding to the flavour-Lagrangian \reff{Lflavour} in the gauge $ U_0= \bone $ is~\cite{Fab02, Fab02b}
\be
{\mathcal T}_{t,t+1} =  \exp  \left(   {\hat q_{-} } \, N_t  \, {\hat q_{+} }   \, \right)^{\dagger}
\exp(2\mu \,  {\hat n}_B)  \exp  \left(   {\hat q_{-} } \, N_{t +1} \, {\hat q_{+} }   \, \right)\,.
\ee
In the above equation   $N_t$ is a matrix which depends on the time of the blocks only because it depends on the gauge link variables
\be
N_t := N[U(t)] \,,
\ee
and $\mu$ is the chemical potential
\be
{\hat n}_B = 2^3 \, \sum_{{\bf y}} \left({\hat q}_{+}^{\dagger} {\hat q}_{+} -  {\hat q}_{-}^{\dagger} {\hat q}_{-} \right)_{{\bf y}}
\ee
that we omitted for simplicity in the Lagrangian.
By keeping into account the spatial volume factors
\be
 {\hat q_{-} } \, N_t  \, {\hat q_{+} } = 64\, \sum_{{\bf y}',\, {\bf y}}  \,  (\hat q_{-} )_{{\bf y}'} ( N_t )_{{\bf y}' {\bf y}} (\hat q_{+})_{\bf y} 
\ee
\be
N_{{\bf y}' {\bf y}} =  -2 \, \Big\{   m \,  (\gamma_0  \otimes \bone)  \, \bone_{{\bf y}' {\bf y}}
  +{ \sum_{k=1}^3}  (\gamma_0  \gamma_k   \otimes \bone) \left[P^{(-)}_k \nabla_k^{(+)}+ P^{(+)}_k \nabla_k^{(-)}\right]_{{\bf y}' {\bf y}}  \vphantom{\sum_{j=0}^3} \Big\} \label{Nstandard}
\ee
where
\begin{equation}
P_k^{(\pm)} = { \frac{1}{2} } ( \bone \otimes \bone \pm \gamma_k \gamma_5 \otimes t_5 t_k ) \, 
\end{equation}
and
\be
\bone_{{\bf y}' {\bf y}} : = \frac{1}{8}\, \delta_{{\bf y}' {\bf y}} \, , \qquad (T_k)^{(\pm)}_{{\bf y}' {\bf y}} := \frac{1}{8}\, \delta_{{\bf y}'\pm \hat{k}, {\bf y}}
\ee
enter in the definitions of $\nabla_k^{(\pm)}$.

Notice that
\be
q_{\pm}^{\dagger} N q_{\pm} =0\,.
\ee

\subsection{The spin basis}

For the sake of later comparison we report the regularization of a Lagrangian in the spin basis.
The gauge fields on the hypercubic lattice are denoted by $u_{\mu}(x)$ and transform according to
\be
u_{\mu}(x) \, \to \,   g_x \, u_{\mu}(x)  \, g^{\dagger}_{x+{\hat \mu}}\,.
\ee
The Lagrangian  in the spin basis  is
\be
{\mathcal L}_{\psi}(u) : = m \, {\overline \psi}_x\psi_x+ \frac{1}{2} \sum_{\mu=0}^3 \alpha_{x\mu}  
\left[  \, {\overline \psi}_x \, u_{\mu}(x)\psi_{x+ {\hat \mu}}   -  {\overline \psi}_{x +{\mu}}  u^{\dagger}_{\mu}(x) \psi_x \right]  
 \label{Lspin}
\ee
where  the signs $\alpha_{x \mu}$ are defined for $\mu=0,\dots,3$ by
\be
\alpha_{x \mu}  := (-1)^{x_0+ \cdots + x_{\mu-1}} \,.
\ee
There is no direct way of  identifying forward and backward movers. This is  the difficulty encountered in the construction of a transfer matrix in operator form for this Lagrangian. 
 Indeed, as far as we know, such a construction has been achieved only after a reduction of the Lagrangian itself,  in which the fermion fields and their conjugates live on odd and, respectively, even  sites \cite{STW}.

At the classical level, however, the fields in the spin and flavour basis  are related according to
\begin{align}
q_{y}^{\alpha a} = &\, \frac{1}{8} \, \sum_\eta \Gamma_{\eta; \alpha a} \,\psi_{2 y + \eta} \label{defq}
 \\
\bar{q}_{y}^{\,a \alpha } = &\, \frac{1}{8} \, \sum_\eta \bar{\psi}_{2 y + \eta} \,\Gamma_{\eta; a \alpha }^\dagger  \label{def}
\end{align}
where 
\be
\Gamma_\eta :=  \gamma_0^{\eta_0} \gamma_1^{\eta_1} \gamma_2^{\eta_2} \gamma_3^{\eta_3} \label{64} \,.
\ee
The matrices $\Gamma$ satisfy the relations
\begin{align}
\frac{1}{4} \tr \left(\Gamma_\eta^\dagger \Gamma_{\eta'} \right) = & \, \delta_{\eta \eta'} \label{65} \\
\frac{1}{4} \sum_\eta \Gamma_{\eta: b\beta}^\dagger \Gamma_{\eta: \alpha a}  = & \, \delta_{b a} \delta_{\beta \alpha} \label{66}
\end{align}
that allow us to invert Eqs.~\reff{def}
\begin{align}
 \psi_{2 y+ \eta} = &\, 2 \tr \left(\Gamma^\dagger_\eta q_y \right)\\
\bar{\psi}_{2 y+ \eta}  = &\, 2 \tr \left(\bar{q}_y \Gamma_\eta  \right)\,.
\end{align}
We will use these relationships in order to derive an action and a transfer matrix in the spin basis from those in the flavour basis.

\section{Transformation of the Lagrangian}\label{tl}

In this Section we express the Lagrangian \reff{Lflavour} in the spin basis using the transformations \reff{def}
\be
\sum_x\,  {\mathcal L}'_\psi(U) := 2^4\, \sum_y {\mathcal L}_q(U)  \, .
\ee
While in the absence of gauge interaction  ${\mathcal L}'_\psi$ coincides with  ${\mathcal L}_{\psi}$, reported in~\reff{Lspin}, we shall see that this does not occur, in general, in the presence of gauge fields. 

The mass term of the action is proportional to
\be
 2^4\,  \sum_y \bar{q}_y q_y =
 \frac{1}{4} \, \sum_y \sum_\eta  \sum_{\eta'} \bar{\psi}_{2y+\eta} \tr \left(\Gamma^\dagger_{\eta '} \Gamma_\eta\right)  \psi_{2y + \eta'}
%  16\,  \sum_y \sum_{a,  \alpha} \bar{q}_y^{a \alpha} q_y^{\alpha a} =
% 4\, \sum_y  \sum_\eta  \sum_{a,  \alpha}\bar{q}_y^{a \alpha} \Gamma_{\eta: \alpha a}  \sum_{b, \beta} \Gamma^\dagger_{\eta: b \beta}  q_y^{\beta b} 
= \sum_x \bar{\psi}_x \psi_x  \,.
 \ee
In order to derive the kinetic term we shall use the relations
\begin{align}
\sum_\alpha \gamma_\mu^{\alpha' \alpha} \Gamma_{\eta:\alpha a} =  & \, \delta_{0 \eta_\mu} \alpha_{\eta \mu} \Gamma_{\eta+\hat{\mu}:\alpha' a} +  \delta_{1\eta_\mu} \alpha_{\eta \mu} \Gamma_{\eta-\hat{\mu}:\alpha' a}\label{70}\\
\sum_{\alpha,a} \gamma_5^{\alpha' \alpha}(t_5 t_\mu)^{a' a}  \Gamma_{\eta:\alpha a} =  & \, - \delta_{0 \eta_\mu} \alpha_{\eta \mu} \Gamma_{\eta+\hat{\mu}:\alpha' a'} +  \delta_{1\eta_\mu} \alpha_{\eta \mu} \Gamma_{\eta-\hat{\mu}:\alpha' a'}\label{71}
\end{align}

From the definition~\reff{64} soon follow the relation~\reff{70} and
\be
\Gamma_\eta \gamma_\mu = (-1)^{ \eta_0 + \eta_1 + \eta_2 + \eta_3 } (-1)^{\eta_\mu} \gamma_\mu \Gamma_\eta 
\ee
so that
\be
\Gamma_\eta \gamma_5 = (-1)^{ \eta_0 + \eta_1 + \eta_2 + \eta_3 } \, \gamma_5 \Gamma_\eta 
\ee
and therefore
\be
\sum_{\alpha,a} \gamma_5^{\alpha' \alpha}(t_5 t_\mu)^{a' a}  \Gamma_{\eta:\alpha a} =  (\gamma_5 \Gamma_\eta \gamma_\mu \gamma_5)_{\alpha' a'} = - (\gamma_5 \Gamma_\eta \gamma_5 \gamma_\mu)_{\alpha' a'} = - (-1)^{\eta_\mu} (\gamma_\mu \Gamma_\eta)_{\alpha' a'} \label{rel32}
\ee
which together with~\reff{70} implies the relation~\reff{71}.

The kinetic term is proportional to
\begin{multline}
\frac{16}{4} \sum_y \sum_\mu \left\{ \bar{q}_y (\gamma_\mu \otimes 1) \left[ U_\mu(y)\, q_{y+\hat{\mu}} - U_\mu^\dagger(y-\hat{\mu})\, q_{y-\hat{\mu}} \right] \right.\\
\left. - \bar{q}_y (\gamma_5 \otimes t_5)\,t_\mu \left[ U_\mu(y)\, q_{y+\hat{\mu}} + U_\mu^\dagger(y-\hat{\mu})\, q_{y-\hat{\mu}} - 2q_y\right] \right\}
\end{multline}
that is
\begin{multline}
\frac{1}{16} \sum_y \sum_\mu \sum_{\eta,\eta'} \sum_{\alpha,\alpha',a,a'} \bar{\psi}_{2y+\eta'} \Gamma^\dagger_{\eta': a' \alpha'}
 \left[ U_\mu(y) \left( \gamma_\mu^{\alpha'\alpha} \delta^{a' a} - \gamma_5^{\alpha'\alpha} (t_5 t_\mu)^{a' a} \right) \Gamma_{\eta:\alpha a} \psi_{2y +2\hat{\mu}+ \eta} \right.  \\
 -\left. U_\mu^\dagger(y-\hat{\mu}) \left( \gamma_\mu^{\alpha'\alpha} \delta^{a' a} + \gamma_5^{\alpha'\alpha} (t_5 t_\mu)^{a' a} \right) \Gamma_{\eta:\alpha a} \psi_{2y -2\hat{\mu}+ \eta} + 2  \gamma_5^{\alpha'\alpha} (t_5 t_\mu)^{a' a} \Gamma_{\eta:\alpha a} \psi_{2y + \eta} \right] 
\end{multline}
which is because of \reff{70} and \reff{71}
\begin{multline}
\frac{1}{8} \sum_y \sum_\mu \sum_{\eta,\eta'}  \sum_{\alpha,\alpha',a,a'}  \bar{\psi}_{2y+\eta'} \Gamma^\dagger_{\eta': a' \alpha'} \alpha_{\eta \mu} \left[ U_\mu(y) \delta_{0 \eta_\mu} \Gamma_{\eta+\hat{\mu}:\alpha' a'}  \psi_{2y +2\hat{\mu}+ \eta}   \right. \\
\left.  - U_\mu^\dagger(y-\hat{\mu}) \delta_{1 \eta_\mu} \Gamma_{\eta-\hat{\mu}:\alpha' a'} \psi_{2y -2\hat{\mu}+ \eta} + (- \delta_{0 \eta_\mu}  \Gamma_{\eta+\hat{\mu}:\alpha' a'} + \delta_{1 \eta_\mu} \Gamma_{\eta-\hat{\mu}:\alpha' a'} ) \psi_{2y + \eta} \right] 
\end{multline}
and performing the trace on spinor and flavour indices~\reff{66}
\begin{multline}
\frac{1}{2}\sum_y \sum_\mu \sum_{\eta,\eta'}  \bar{\psi}_{2y+\eta'}  \alpha_{\eta \mu}
\left[ U_\mu(y) \delta_{0 \eta_\mu} \delta_{\eta',\eta+\hat{\mu}}  \psi_{2y +2\hat{\mu}+ \eta} 
  \right. \\
\left.  - U_\mu^\dagger(y-\hat{\mu}) \delta_{1 \eta_\mu} \delta_{\eta',\eta-\hat{\mu}} \psi_{2y -2\hat{\mu}+ \eta} + (- \delta_{0 \eta_\mu}  \delta_{\eta',\eta+\hat{\mu}} + \delta_{1 \eta_\mu} \delta_{\eta',\eta-\hat{\mu}} ) \psi_{2y + \eta} \right] 
\end{multline}
and performing the sum over $\eta'$
\begin{multline}
\frac{1}{2}\sum_y \sum_{\eta} \sum_\mu   \alpha_{\eta \mu}
\left[ \, \delta_{0 \eta_\mu}  \bar{\psi}_{2y+\eta+\hat{\mu} } \,U_\mu(y) \,  \psi_{2(y +\hat{\mu})+ \eta } 
+\ \delta_{1 \eta_\mu} \bar{\psi}_{2y+\eta-\hat{\mu}} \, \psi_{2y + \eta} 
     \right. \\ 
\left.- \delta_{1 \eta_\mu}  \bar{\psi}_{2y+\eta-\hat{\mu}} \,U_\mu^\dagger(y-\hat{\mu}) 
\, \psi_{2(y-\hat{\mu}) + \eta } -\delta_{0 \eta_\mu} \bar{\psi}_{2y+\eta+\hat{\mu}}  \,  \psi_{2y + \eta} \right] 
\,.
\end{multline}

Remark that if we increase the component $x_\mu$ of a site $x$ we jump on block different from  that of $x$ if $x_\mu$ is odd. This is the case when $x=2y+\eta+\hat{\mu}$ and $\eta_\mu=0$, but not when $x=2y+\eta-\hat{\mu}$ and $\eta_\mu=1$. Similarly, if we decrease $x_\mu$ we jump on a different block only when $x_\mu$ is even. This is the case when $x=2y+\eta-\hat{\mu}$ and $\eta_\mu=1$, but not when $x=2y+\eta+\hat{\mu}$ and $\eta_\mu=0$. 
And that, if $x = 2y +\eta$ then
\be
\alpha_{\eta \mu} = \alpha_{x \mu}\, .
\ee
Then the kinetic term has the  form as that of ${\mathcal L}_{\psi}(u')$
%\be
%\frac{1}{2}\sum_x\, \alpha_{x \mu}\, \left[\bar{\psi}_x U_\mu'(x) \psi_{x+\hat{\mu}} - \bar{\psi}_{x+\hat{\mu}} (U_\mu')^\dagger(x)\psi_x\right]
%\ee
where
\be
u_{\mu}'( x) = \begin{cases}
U_{\mu}(y) &  \hbox{for\, }  x = 2y + \eta \, \hbox{ and } \eta_\mu =1\\ 
\bone  &  \hbox{elsewhere }  
\end{cases} \label{39}
\ee
that is  the gauge field couples only sites which belong to different blocks.

In conclusion
\be
{\mathcal L}'_\psi( u') = m \, {\overline \psi}_x\psi_x+ \frac{1}{2} \sum_{\mu=0}^3 \alpha_{x\mu}  
\left[  \, {\overline \psi}_x \, u_{\mu}'(x)\psi_{x+ {\hat \mu}}   -  {\overline \psi}_{x +{\mu}}  u'^{\dagger}_{\mu}(x) \psi_x \right]  \,.\label{L'}
\ee
We have the constraint, however, that the fermion fields within a block should all transform in the same way under gauge transformations.
One might think that we could relax this constraint by a different   transformation from the spin to the flavour basis
\begin{align}
q_{y}^{\alpha a} = &\, \frac{1}{8} \, \sum_\eta \Gamma_{\eta; \alpha a} \,{\mathcal C}_{2y +\eta}\psi_{2 y + \eta} 
\nonumber\\
\bar{q}_{y}^{\,a \alpha } = &\, \frac{1}{8} \, \sum_\eta \bar{\psi}_{2 y + \eta} \,{\mathcal C}_{2y + \eta}^{\dagger}\Gamma_{\eta; \alpha a}^\dagger  \, .
\end{align}
Such a generalization, however, is only apparent, because the curvature for the plaquettes with all the vertices within one and the same  block vanishes. Indeed, such a generalization, as the particular ones chosen for example in~\cite[Eq.~(35)]{Kluber83}, \cite[Eq.~(56)]{Fab02b} is a pure-gauge transformation of~\reff{def}.

We conclude that, in the presence of a {\em generic} gauge-field configuration, the Lagrangian in the spin basis ${\mathcal L}'_\psi( u)$ and that in the flavor basis ${\mathcal L}_q( U)$ are not equivalent.

The transformed Lagrangian ${\mathcal L}'_\psi(u')$ could also be regarded, in the spirit of the previous quoted attempt~\cite{STW}, as a modification of 
${\mathcal L}_{\psi}(u)$, defined in~\reff{Lspin},  for  which a transfer matrix can be constructed.

The above construction refers to the case of vanishing chemical potential. Its inclusion is, however, straightforward~\cite{Fab02b}. We only note that, at variance with respect to the coupling with gauge fields, the chemical potential can be attached to all links in the transformed Lagrangian ${\mathcal L}'_\psi(u')$, provided its value be half the one in the flavour basis.

\section{Transformation of  transfer matrix and coherent states}\label{ttm}

As a first step we must transform  creation-annihilation operators from the flavour to the  spin basis. To this end we must determine the expressions of the fields $q_{\pm}$ in the spin basis
\be
(q_{+})_y= P_+ \frac{1}{8}\sum_{\eta}\Gamma_{\eta} \psi_{2 y+\eta} \, \qquad 
(q_{-}^\dagger)_y= P_- \frac{1}{8}\sum_{\eta}\Gamma_{\eta} \psi_{2 y+\eta} \,.
\ee
Using the relation~\reff{rel32}, 
%\be
%\gamma_0 \gamma_5 \otimes t_5t_0 \Gamma_{\eta} = - \gamma_0 \gamma_5 \Gamma_{\eta}\gamma_5 \gamma_0 
%= - (-1)^{\eta_0}\Gamma_{\eta}
%\ee
we find
\be
P_{+}\Gamma_{\eta}= \delta_{0 \eta_0} \Gamma_{\eta}\, , \qquad \, P_{-}\Gamma_{\eta}= \delta_{1 \eta_0} \Gamma_{\eta}\, \label{PGamma}
\ee
and similar relations hold for $\Gamma^\dagger$.
%In view of the special role played by the temporal component of $\eta$, we adopt the notation
%\be
%  \psi_{2y+\eta} =  \psi_{2 y+{\vec \eta}, \eta_0}  \,,  
% \ee

We therefore have 
%$\boldsymbol{\eta}$
\be
(q_{+})_y= \frac{1}{8}\, \sum_{\eta} \delta_{0 \eta_0} \Gamma_\eta\psi_{2 y+\eta} \, , \qquad
(q_{-}^\dagger)_y= \frac{1}{8}\, \sum_{\eta} \delta_{1 \eta_0} \Gamma_\eta\psi_{2 y+\eta} \, .
\ee
Next we define the operators corresponding to the $\psi$-fields according to
\be
({\hat q}_{+})_y= \frac{1}{8}\, \sum_{\boldsymbol{\eta}} \delta_{0 \eta_0} \Gamma_\eta {\hat \psi}_{2 y+\eta} \, , \qquad
({\hat q}_{-}^\dagger)_y= \frac{1}{8}\, \sum_{\eta} \delta_{1 \eta_0} \Gamma_\eta {\hat \psi}_{2 y+\eta} \label{52}
\ee
and assume that
\be
\{ {\hat \psi}_{2{\bf y}'+\eta'}^{\dagger}, {\hat \psi}_{2{\bf y}+\eta} \} =  2\, \delta_{{\bf y}' {\bf y}} \delta_{\eta' \eta} \,. \label{quantum number}
\ee
This is obviously consistent with the second set of  equations in~\reff{comm}. Consistency with  the first set requires that
\begin{align}
 \frac{1}{64} \, \sum_{\eta, \eta'} \delta_{\sigma \eta_0} \delta_{\tau \eta'_0} \Gamma_{\eta':b \beta}^{\dagger} \Gamma_{\eta:\alpha a} 
\{ {\hat \psi}_{2 \mathbf{y}'+ \eta'}^{\dagger}, {\hat \psi}_{2 \mathbf{y} + \eta} \} =  & 
 \frac{1}{32} \, \delta_{{\bf y}' {\bf y}}  \delta_{\sigma \tau} \, \sum_{\eta} \delta_{\sigma \eta_0} \Gamma_{\eta:b \beta}^{\dagger} \Gamma_{\eta:\alpha a} \\
= & \frac{1}{8} \, \delta_{{\bf y}' {\bf y}} \delta_{\sigma \tau} P_\pm ^{\alpha a, \beta b  }  \, 
\end{align}
where $\sigma = 0, 1$ respectively when the index of the projector is $+$ or $-$. The second equality follows from the equations
\be
\sum_{a', \alpha'} P_\pm^{\alpha a , \alpha' a' }  \, \frac{1}{4} \sum_\eta \Gamma_{\eta: b\beta}^\dagger \Gamma_{\eta: \alpha' a'}  =   
\frac{1}{4} \sum_\eta  \delta_{\sigma \eta_0} \Gamma_{\eta: b\beta}^\dagger \Gamma_{\eta: \alpha a}  =
P_\pm ^{\alpha a, \beta b  } 
\ee
that can be proven using ~\reff{66} and \reff{PGamma}.

Some comments about our results are in order. We see that the temporal component $\eta_0$ of the fields in the spinor basis corresponds to the $\pm$ projection  of the field in the flavor basis. The 8 Dirac-taste degrees of freedom of particles/antiparticles   are spread on the 8 sites of the even/odd  time slice in the corresponding block. In this connection, looking at Eq.\reff{quantum number},  $\eta_0$ can be regarded as a quantum number. But this quantum number changes when time increases by one unit in the original lattice, so that, unlike the $q_{\pm}$ projections,   the fields $\psi_{2 \mathbf{y} + \eta}$ with $\eta_0$ respectively 1 or 0 cannot be identified as forward/backward movers. Changing time we change a particle into the hole of an antiparticle.

\subsection{Transfer matrix}

We first   transform  the baryon number
\begin{align}
{\hat n}_B = 2^3 \, \sum_{{\bf y}} \left({\hat q}_{+}^{\dagger} {\hat q}_{+} -  {\hat q}_{-}^{\dagger} {\hat q}_{-} \right)_{{\bf y}}
= & \, \frac{1}{2} \, \sum_{{\bf y}, \eta}  \left[ \left( {\hat \psi}^{\dagger} {\hat \psi}  \right)_{2 {\bf y} +  \eta}   \delta_{0 \eta_0} -  \left( {\hat \psi}\, {\hat \psi}^{\dagger}  \right)_{2 {\bf y} +  \eta}   \delta_{1 \eta_0} ) \right]\\
= & \, \frac{1}{2} \, \sum_{{\bf x}}  \left[ \left( {\hat \psi}^{\dagger} {\hat \psi}  \right)_{{\bf x} 0}    -  \left( {\hat \psi}^{\dagger}  {\hat \psi} \right)_{ {\bf x} 1}  \right]  \,
\end{align}
where we re-label the operators $\hat{\psi}$ with the spatial coordinates
\be
{\bf x} = 2\, {\bf y} + \boldsymbol{\eta}
\ee
and $\eta_0$ and made the identifications
\be
\hat{\psi}_{{\bf x} 0} := \hat{\psi}_{2 {\bf y} + (0, \boldsymbol{\eta})} \, , \qquad \hat{\psi}_{{\bf x} 1} := \hat{\psi}_{2 {\bf y} + (1, \boldsymbol{\eta})} ^\dagger
\ee
in agreement with the relations~\reff{52} which show that when $\eta_0=1$ the operator ${\hat \psi}_{2 {\bf y} + \eta}$ is a creation operator.

In this notation the commutation relations~\reff{quantum number} become
\be
\{ {\hat \psi}_{{\bf x}' \eta_0'}^{\dagger}, {\hat \psi}_{{\bf x} \eta_0} \} =  2\, \delta_{{\bf x}' {\bf x}} \delta_{\eta_0' \eta_0} \,. \label{acr}
\ee
Next we must  determine a matrix $N_t' $ such that
\begin{align}
64\, \sum_{ {\bf y}',\, {\bf y}} (\hat q_{-})_{{\bf y}'}  (N_t)_{ {\bf y}' {\bf y}}  \,(\hat q_{+})_{\bf y}  =    &  \sum_{{\bf y}',\,  {\bf y}} \sum_{\eta',\,  \eta}
{\hat \psi}_{2 {\bf y}' + \eta'}^{\dagger}\,  \tr \left(  \Gamma_{\eta'}^\dagger (N_t)_{ {\bf y}' {\bf y}} P_+
 \Gamma_{\eta} \right) \,{\hat \psi}_{2{\bf  y} + \eta} \\
 =  &   \sum_{{\bf y}',\,  {\bf y}} \sum_{\eta',\,  \eta}
{\hat \psi}_{2 {\bf y}' + \eta'}^{\dagger}\,  (N'_t)_{{\bf y}' \eta', {\bf y}  \, \eta}   \, {\hat \psi}_{2{\bf  y} + \eta} \,.
\end{align}
In the above equation color taste and Dirac indices have been omitted. We observe that 
\begin{align}
(\gamma_0\gamma_k \otimes \bone)  \, P_k^{(\pm)} P_{+} \Gamma_{\eta} & = \frac{1}{2} \delta_{0 \eta_0}
\left[ (\gamma_0 \gamma_k \otimes \bone) \pm (\gamma_0 \gamma_5 \otimes t_5 t_k) \right] \Gamma_{\eta} \\
&= \delta_{0 \eta_0}\alpha_{\eta k} \, \frac{ 1 \mp ( -1)^{\eta_k} }{2} \,
( \delta_{0\eta_k} \Gamma_{\eta+{\hat 0} + {\hat k} }  + \delta_{1 \eta_k} \Gamma_{\eta+{\hat 0} - {\hat k} } )
\end{align}
and
\be
\Gamma_{\eta'}^{\dagger} (\gamma_0 \otimes \bone) \, P_{+} \Gamma_{\eta} = \delta_{0 \eta_0} \delta_{1 \eta_0'} \Gamma_{\eta'}^{\dagger}\Gamma_{\eta }\,,
\ee
so that
\be
\tr \left[ \Gamma_{\eta'}^{\dagger} (\gamma_0\gamma_k \otimes \bone)  \, P_k^{(\pm)} P_{+} \Gamma_{\eta}  \right]
 =   4\, \delta_{0 \eta_0} \delta_{1 \eta_0'}\alpha_{\eta k} \, \frac{ 1 \mp ( -1)^{\eta_k}} {2} \,
( \delta_{0 \eta_k} \delta_{\boldsymbol{\eta}', \boldsymbol{\eta} + {\hat k}} + \delta_{1 \eta_k} \delta_{\boldsymbol{\eta}', \boldsymbol{\eta} - {\hat k}})
\ee
and
\be
\tr \left[ \Gamma_{\eta'}^{\dagger} (\gamma_0 \otimes \bone)  \, P_{+} \Gamma_{\eta}  \right]
=4 \, \delta_{0 \eta_0} \delta_{1 \eta_0'}  \delta_{\boldsymbol{\eta}' \boldsymbol{\eta}}  \,.
\ee
Finally we get the transformed $N$-matrix
\begin{align}
(N' )_{{\bf y}' \eta', {\bf y}  \, \eta} = & - 8 \, \delta_{0 \eta_0} \, \delta_{1 \eta_0'}  \left[ m \,  \delta_{ \boldsymbol{\eta}' \boldsymbol{\eta}}  \bone_{{\bf y}' {\bf y}}
+ \sum_{\mu=1}^3 \alpha_{\eta \mu} \left( \delta_{0\eta_\mu} \delta_{ \boldsymbol{\eta}', \boldsymbol{\eta}+{\hat \mu}} \nabla_\mu^{(+)}
+ \delta_{1\eta_\mu} \delta_{ \boldsymbol{\eta}', \boldsymbol{\eta}-{\hat \mu}} \nabla_\mu^{(-)}
\right)_{{\bf y}' {\bf y}} \right] 
\nonumber\\
= & \, - \, \delta_{0 \eta_0} \, \delta_{1 \eta_0'}  \left\{  m \,  \delta_{ \boldsymbol{\eta}' \boldsymbol{\eta}}  \delta_{{\bf y}', {\bf y}} 
 + \frac{1}{2}   \sum_{\mu=1}^3 \alpha_{\eta \mu} \left[ \vphantom{U_\mu^{\dagger}} \left(- \delta_{0\eta_\mu} \delta_{ \boldsymbol{\eta}', \boldsymbol{\eta}+{\hat \mu}} + \delta_{1\eta_\mu} \delta_{ \boldsymbol{\eta}', \boldsymbol{\eta}-{\hat \mu}} \right) \delta_{{\bf y}' {\bf y}}   \right.
 \right. 
 \nonumber\\
& \quad +  \left. \left.     \delta_{0\eta_\mu} \delta_{ \boldsymbol{\eta}', \boldsymbol{\eta}+{\hat \mu}}  \, U_\mu({\bf y}') \delta_{{\bf y}, {\bf y}'+{\hat \mu}} 
- \delta_{1\eta_\mu} \delta_{ \boldsymbol{\eta}', \boldsymbol{\eta}-{\hat \mu}}  U_\mu^{\dagger}({\bf y}) \delta_{{\bf y}, {\bf y}'-{\hat \mu}} 
  \right] \vphantom{\sum_{j=1}^3} \right\} \,.
\end{align}
Notice that the terms that involve the gauge variables refer to sites belonging to different blocks, while in the other terms the sites belong to the same blocks.
The same operator can be re-labelled by using the coordinates $\bf x$ and $\eta_0$, then
\begin{multline}
(N' )_{{\bf x}' \eta'_0, {\bf x}  \, \eta_0} =  \, - \, \delta_{0 \eta_0} \, \delta_{1 \eta_0'}   \\
\left\{  m \,  \delta_{{\bf x}' {\bf x}} + \frac{1}{2}   \sum_{\mu=1}^3  \alpha_{{\bf x} \mu}   \left[  \delta_{{\bf x}', {\bf x} - \hat{\mu}}  u'_\mu ({\bf x}') -  \delta_{{\bf x}', {\bf x} + \hat{\mu}}  u'^{\dagger}_{\mu} ({\bf x}) \right]  \right\} \label{N'}
\end{multline}
where the values $\eta_\mu = 0,1$  simply control the presence of the gauge field according to the definition of $u'$ given in~\reff{39}.

In conclusion
\be
{\hat q_{-} } \, N_t  \, {\hat q_{+} } \, =  \, {\hat \psi_{1} } \, N'_t  \, {\hat \psi_{0} } \, .
\ee
It should not be necessary to repeat that the expression of the transfer matrix so obtained is positive definite and performs time translations by two lattice spacings.

%@@@@@@@@@@@@@@@@@@@@@@@@@@@@@@@
%
%\begin{align}
%\nonumber
%q_{y}^{\alpha a} = &\, \frac{1}{8} \, \sum_\eta \Gamma_{\eta; \alpha a} \,{\mathcal C}_{2y,\eta}\psi_{2 y + \eta} \\
%\bar{q}_{y}^{\,a \alpha } = &\, \frac{1}{8} \, \sum_\eta \bar{\psi}_{2 y + \eta} \,{\mathcal C}_{2y,\eta}^{\dagger}\Gamma_{\eta; \alpha a}^\dagger  
%\end{align}
%\be
%{\mathcal C}_{2y,\eta}'= g_{2y}\, {\mathcal C}_{2y,\eta} \, g_{2y+\eta}^{\dagger}.
%\ee
%\be
%{\mathcal C}_{2y,\eta- {\hat \mu}}^{\dagger} {\mathcal C}_{2y,\eta}\,, \,\,\, {\mathcal C}_{2(y+{\mu}),\eta}^{\dagger}U_\mu(y) \, {\mathcal C}_{2(y+{\hat \mu}), \eta}
%\ee
%
%@@@@@@@@@@@@@@@@@@@@@@@@@@@@@@@
%
%Altri lavori potenzialmente utili: un altro di Smit~\cite{BHS}, fermioni di Kaplan~\cite{Shamir}, domain wall dell'amico tuo americano~\cite{Negele}, formulazione Hamiltoniana nonrelativistica~\cite{El-Khadra}, e mi pare che anche questo lo conosci~\cite{Mitrjushkin}

%\bibliography{TransferMatrixFermions}{}

\begin{thebibliography}{10}

\bibitem{Wilson74}
K.~G.~Wilson, 
\emph{Confinement of quarks\/}, 
Phys.\ Rev.\ D \textbf{10} (1974) 2445.

\bibitem{Wilson75}
K.~G.~Wilson, 
\emph{Quarks and strings on a lattice\/},
Erice Lectures 1975, ed. A.~Zichichi, Plenum Press Corporation, New York, 1977.

\bibitem{KS}
J.~Kogut and L.~Susskind,
\emph{Hamiltonian formulation of Wilson's lattice gauge theories\/}, 
Phys.\ Rev.\ D \textbf{11} (1975) 395.

\bibitem{BKS}
T.~Banks, J.~Kogut, and L.~Susskind, 
\emph{Strong coupling calculations of lattice gauge theories: (1+1) dimensional exercises\/}, 
Phys.\ Rev.\ D \textbf{13} (1976), 1043.

\bibitem{SS76}
L.~Susskind, 
\emph{Weak and electromagnetic interactions at high-energy\/},
Les Houches Lectures, Session XXIX, eds. R.~Balian and C.~H.~Llewellyn Smith, North Holland, 1976.

\bibitem{Susskind}
L.~Susskind, 
\emph{Lattice fermions\/}, 
Phys.\ Rev.\ D \textbf{16} (1977) 3031.

\bibitem{MM}
I.~Montvay and G.~M\"unster, 
\emph{Quantum fields on a lattice\/}, 
Cambridge Monographs on Mathematical Physics, Cambridge University Press, 1994.


\bibitem{Kluber81}
H.~Kluberg-Stern, A.~Morel, O.~Napoly, and B.~Petersen, 
\emph{Spontaneous symmetry breaking for U(N) gauge theory on a lattice\/}, 
Nucl. Phys. B \ \textbf{190 [FS3]} (1981) 504.

\bibitem{Kluber83}
H.~Kluberg-Stern, A.~Morel, O.~Napoly, and B.~Petersen, 
\emph{Flavours of Lagrangian Susskind fermions\/}, 
Nucl.\ Phys.\ B \textbf{220 [FS8]} (1983) 447.

\bibitem{Gliozzi}
F.~Gliozzi, 
\emph{Spinor algebra of one component lattice fermions\/},
Nucl.\ Phys.\ B \textbf{204} (1982) 419.

\bibitem{STW}
H.~S. Sharatchandra, H.~J. Thun, and P.~Weisz, 
\emph{Susskind fermions on a euclidean lattice\/}, 
Nucl.\ Phys.\ B \textbf{192} (1981) 205.

\bibitem{Vandendoel}
C.~P.~van den Doel and J.~Smit, 
{\em Dynamical symmetry breaking in two flavor SU(N) and SO(N) lattice gauge theories\/},
Nucl.\ Phys.\ {\bf B 228} (1983) 122.

\bibitem{Golterman}
M~Golterman and J.~Smit, 
\emph{Selfenergy and flavour interpretation of staggered fermions\/}, 
Nucl.\ Phys.\ B \textbf{245} (1984) 61.

\bibitem{Fab02}
P.~Palumbo, 
{\em The transfer matrix with Kogut-Susskind fermions\/},
Phys.\ Rev.\ D {\bf 66} (2002) 077503  [arXiv:hep-lat/0208005]; 
Erratum-ibid. {\bf 73} (2006) 119902. 
%%CITATION = HEP-LAT 0208005;%%
 
\bibitem{LuscherTM}
M.~L\"{u}scher, 
\emph{Construction of a selfadjoint, strictly positive transfer matrix for euclidean lattice gauge theories\/}, 
Commun.\ Math.\ Phys.\  {\bf 54} (1977) 283.
%%CITATION = CMPHA,54,283;%%

\bibitem{Creutz77}
M.~Creutz, 
\emph{Gauge fixing, the transfer matrix, and confinement on a lattice\/}, 
Phys.\ Rev.\ D \textbf{15} (1977) 1128.

\bibitem{Creutz87}
M.~Creutz, 
\emph{Species doubling and transfer matrices for fermionic fields\/},
Phys.\ Rev.\ D \textbf{35} (1987) 1460.

\bibitem{Smit}
J.~Smit, 
\emph{Transfer operator and lattice fermions\/}, 
Nucl.\ Phys.\ Proc.\ Suppl. \textbf{20} (1991) 545.

\bibitem{Creutz99}
M.~Creutz,  
\emph{Transfer matrices and lattice fermions at finite density\/},
  Foundations of Physics \textbf{30} (2000) 487.
 
\bibitem{MP}
P.~Menotti and A.~Pelissetto,
{\em General proof of Osterwalder-Schrader positivity for the Wilson action\/},
Commun.\ Math.\ Phys.\ {\bf 113} (1987) 369.
%%CITATION = CMPHA,113,369;%%

\bibitem{Banks}
T.~Banks and A.~Zaks, \emph{Chiral analog gauge theories on the lattice\/}, 
Nucl.\ Phys.\ B \textbf{206} (1982) 23.

\bibitem{CLP}
S.~Caracciolo, V.~Laliena and F.~Palumbo,
{\em Composite boson dominance in relativistic field theories\/},
JHEP {\bf 0702} (2007) 034 [arXiv:hep-lat/0611012].
%%CITATION = JHEPA,0702,034;%%

\bibitem{Palu}
F.~Palumbo, {\em Semivariational approach to QCD at finite temperature and baryon density},
Phys. Rev. D {\bf 78} (2008) 014514
\bibitem{CPV}
S.~Caracciolo,  F.~Palumbo and G.~Viola,
{\em Bogoliubov transformations and fermion condensates in lattice field theories\/},
Annals Phys.\  {\bf 324} (2009) 584 [arXiv:0808.1110].
%%CITATION = APNYA,324,584;%%

\bibitem{noi-prd}
S.~Caracciolo and F.~Palumbo, 
{\em Chiral symmetry breaking and quark confinement in the nilpotency expansion of QCD\/},
Phys.\ Rev.\  {\bf D 83} (2011) 114504 [arXiv:1010.0596].
%%CITATION = PHRVA,D83,114504;%%

\bibitem{diquarks}
S.~Caracciolo and F.~Palumbo, 
{\em Diquarks in the nilpotency expansion of QCD and their role at finite chemical potential\/},
Phys.\ Rev.\  {\bf D 85} (2012) 094009 [arXiv:1112.3493].
%%CITATION = PHRVA,D85,094009;%%

\bibitem{Fab02b}
P.~Palumbo, 
{\em The chemical potential in the transfer matrix and in the path integral formulation of QCD on a lattice\/},
Nucl.\ Phys.\ B \textbf{645} (2002) 309.

%
%\bibitem{BHS}
%W.~Bock, J.~E. Hetrick, and J.~Smit, 
%\emph{Fermion production despite fermion number conservation\/}, 
%Nucl.\ Phys.\ B \textbf{437} (1995) 610.
%
%\bibitem{El-Khadra}
%A.~X.~El-Khadra, A.~S. Kronfeld, and P.~B.~Mackenzie, 
%\emph{Massive fermions in lattice gauge theory\/}, 
%Phys.\ Rev.\ D \textbf{55} (1997) 3933.
%
%\bibitem{Shamir}
%V.~Furman and Y.~Shamir, 
%\emph{Axial symmetries in lattice QCD with Kaplan fermions\/}, 
%Nucl.\ Phys.\ B \textbf{439} (1995) 78.
%
%\bibitem{Mitrjushkin}
%V.~K. Mitrjushkin, 
%\emph{Transfer matrix and nonperturbative renormalization of fermionic currents in lattice QCD\/},  
%(2002), [arXiv:hep-lat/0206024v1].
%
%\bibitem{Nagele}
%S.~Syritsyn and J.~W. Negele, 
%\emph{Oscillatory terms in the domain wall transfer matrix}, 
%PoS (LATTICE 2007) (2008) 078.

\end{thebibliography}
%\bibliographystyle{amsplain}

\subsection{Coherent states}

In order to complete our analysis we perform the transformation also on the coherent states. This will enable us to make, as a crosscheck,
the  derivation of the Lagrangian~\reff{L'} starting from the transfer matrix.

Let 
\be
|\alpha, \beta \rangle := \exp \left\{ - \, 2^3 \, \sum_{\bf y} \, \sum_{\gamma, c} \, [ \alpha_{\bf y}^{\gamma c} ( \hat{q}^\dagger_+)_{\bf y}^{c \gamma} +
\beta_{\bf y}^{c \gamma} ( \hat{q}^\dagger_-)_{\bf y}^{ \gamma c} ]\right\} \, | 0 \rangle
\ee
be a coherent state in the flavour basis, where $\alpha_{\bf y}^{\gamma c}$ and $\beta_{\bf y}^{\gamma c}$ are Grassmann variables, such that
\be
(\hat{q}_+)_{\bf y}^{\gamma c}\, |\alpha, \beta \rangle = \alpha_{\bf y}^{\gamma c} \,  |\alpha, \beta \rangle \, , \qquad
(\hat{q}_-)_{\bf y}^{c \gamma}\, |\alpha, \beta \rangle = \beta_{\bf y}^{c \gamma } \,  |\alpha, \beta \rangle 
\ee
Now
\begin{align}
2^3 \, \sum_{\bf y} \, \sum_{\gamma, c} \,  \alpha_{\bf y}^{\gamma c} ( \hat{q}^\dagger_+)_{\bf y}^{c \gamma} =  & 
  \sum_{{\bf y}, \eta} \,  \tr \left( \Gamma_\eta^\dagger  \alpha_{\bf y} \right)  \delta_{0\eta_0} \, \hat{\psi}^\dagger_{2 {\bf y} + \eta} \\
2^3 \, \sum_{\bf y} \, \sum_{\gamma, c}  \beta_{\bf y}^{c \gamma } ( \hat{q}^\dagger_-)_{\bf y}^{\gamma c} =  & 
  \sum_{{\bf y}, \eta} \, \tr \left(  \beta_{\bf y}  \Gamma_\eta  \right)  \delta_{1\eta_0} \, \hat{\psi}_{2 {\bf y} + \eta} 
\end{align}
and therefore, because of the anti-commutation relations~\reff{acr}
\begin{align}
\hat{\psi}_{{\bf x} 0}\, |\alpha, \beta \rangle = \sum_{\eta_0}  \hat{\psi}_{2 {\bf y} +  \eta}\,  \delta_{0\eta_0} \, |\alpha, \beta \rangle = 2 \tr \left( \Gamma_{(0, \boldsymbol{\eta})}^\dagger  \alpha_{\bf y} \right) |\alpha, \beta \rangle \\
\hat{\psi}_{{\bf x} 1}\, |\alpha, \beta \rangle = \sum_{\eta_0}  \hat{\psi}_{2 {\bf y} +  \eta}^\dagger\,  \delta_{1\eta_0} \, |\alpha, \beta \rangle = 2 \tr \left(  \beta_{\bf y} \Gamma_{(1, \boldsymbol{\eta})}  \right) |\alpha, \beta \rangle \, .
\end{align}
This means that we can define
\be
\alpha_{\bf x}' := 2 \tr \left( \Gamma_{(0, \boldsymbol{\eta})}^\dagger  \alpha_{\bf y} \right) \, , \qquad
\beta_{\bf x}' := 2 \tr \left(  \beta_{\bf y} \Gamma_{(1, \boldsymbol{\eta})} \right) 
\ee
and re-write
\be
|\alpha, \beta \rangle = \exp \left[ - \frac{1}{2} \, \sum_{\bf x} \left(  \alpha_{\bf x}' \hat{\psi}^\dagger _{{\bf x} 0} + \beta_{\bf x}' \hat{\psi} _{{\bf x} 1}^\dagger \right) \right] \, | 0 \rangle \, .
\ee
Notice that the Grassmann variables $\alpha, \beta$ and $\alpha'$ as well are defined at even times. The variable $\beta'$ instead, because of the matrix $\Gamma_{(1,  \boldsymbol{\eta})}$ in its definition, must be considered attached at odd times. This is confirmed by the evaluation of the partition function using the transformed transfer matrix and coherent states. After the identifications
\begin{eqnarray}
{\overline \psi}_{2x_0}&&=( \alpha'_{2x_0})^* \,, \,\,\, \psi_{2x_0}=( \beta'_{2x_0+1})^*
\nonumber\\
{\overline \psi}_{2x_0+1}&&= \beta'_{2x_0+3} \,, \,\,\, \psi_{2x_0+1}= \alpha'_{2x_0+2}
\end{eqnarray}
we get  the Lagrangian~\reff{L'}. 

\section{Conclusion}

Numerical simulations with Kogut-Susskind fermions are faster in the spin basis than in the flavor basis. Such calculations are usually performed in the lagrangian formulation, but we are interested in numerical simulations in the framework of the nilpotency expansion, that makes use of the transfer matrix. So we need an expression of the  transfer matrix in the spin basis. In any case the knowledge of a positive definite transfer matrix in the spin basis is {\em per se} interesting being related to the unitarity of the theory.

We found in the literature essentially two formulations of the transfer matrix in the spin basis. In the first one the Lagrangian is reduced by defining fermion fields and their conjugates at the odd, respectively even sites, and  a transfer matrix is constructed that performs time translations by 2 lattice spacings~\cite{STW,Vandendoel}. The fermion determinant even at vanishing chemical potential, is, however, not positive definite, which makes this way less suitable to numerical simulations.

In the second formulation~\cite{STW}, a positive definite transfer matrix, called $T^2$, was defined that also  performs time translations by 2 lattice spacings.  As a consequence the corresponding Fock space must be constructed on blocks. The explicit construction of such Fock space, however, is not given. 

If the Fock space  is associated to a block,  we can get the transfer matrix in the spin basis by a unitary transformation from that in the flavor basis, whose expression, together with the construction of the Fock space, are known. The transfer matrix in the flavor basis is expressed in terms of a matrix $N$ , and the transformed matrix is given in terms of  the matrix $N'$, given explicitly in~\reff{N'}. In order to do numerical simulations in the nilpotency expansion all we need is to replace everywhere in the equations of the nilpotency expansion $N$ by $N'$ and remember that the gauge fields are now defined on blocks.

It would be now natural to compare our result with the expression of the previously derived transfer matrix~\cite{STW}. One might expect that such a comparison should provide the definition of the Fock space in the latter. Unfortunately this is not the case. The transfer matrix of~\cite{STW}  cannot be related to ours in a simple way,  the most remarkable differences being that there is no requirement concerning the gauge variables which remain defined on the links of the original lattice, and creation and annihilation operators appear not only in exponential form but also as powers.

\end{document}